\documentclass[12pt, draftclsnofoot, onecolumn, romanappendices]{IEEEtran}

\usepackage{amsmath}
\usepackage{amsthm}
\usepackage{amsbsy}
\usepackage{amssymb}
\usepackage{bm}
\usepackage[noadjust]{cite}
\usepackage{graphicx}
\usepackage{float}
\usepackage{epsfig}
\usepackage{microtype}
\usepackage{mathtools}
\usepackage{latexsym}
\usepackage{wasysym}
\usepackage[subtle]{savetrees}
\usepackage{placeins}
\usepackage{color}
\usepackage{rotating}
\usepackage[usenames,dvipsnames]{xcolor}
\usepackage{caption}
\usepackage{subcaption}
\usepackage{epstopdf}
\usepackage{cancel}
\usepackage{tikz}
\usetikzlibrary{fit,positioning,arrows.meta, backgrounds}
\usepackage{tkz-euclide}
\usetkzobj{all}
\usepackage{algorithm}
\usepackage{algpseudocode}
\usepackage{overpic}
\usepackage{dashrule}
\usepackage{subdepth}
\usepackage{overpic}
\usepackage{stackengine}
\usepackage{accents}
\usepackage{stmaryrd}
\usepackage{xpatch}

\newtheorem{lemma}{Lemma}

\DeclareMathAlphabet{\mathbit}{OML}{cmr}{bx}{it}
\DeclareMathAlphabet{\mathsf}{OT1}{cmss}{m}{n}
\DeclareMathAlphabet{\mathTXf}{OT1}{cmss}{bx}{it}

\DeclareMathOperator*{\argmax}{argmax}

\theoremstyle{remark}
\newtheorem{remark}{Remark} 
\theoremstyle{example}
\theoremstyle{assumption}
\newcommand{\ubar}[1]{\text{\b{$#1$}}}
\IEEEoverridecommandlockouts
\definecolor{orange}{rgb}{0.8627,0.4314,0.1961}
\definecolor{grey}{rgb}{0.2196,0.2471,0.3176}
\definecolor{blue}{rgb}{0.0275,0.4431,0.5294}
\definecolor{green}{rgb}{0,0.6078,0.4392}
\begin{document}

\bstctlcite{IEEEexample:BSTcontrol}
\title{Exploring the Trade-Off between Privacy and Coordination in mmWave Spectrum Sharing}
\author{
     	\IEEEauthorblockN{
     		Flavio Maschietti, Paul de Kerret, David Gesbert}
     	\IEEEauthorblockA{ 
     		Communication Systems Department, EURECOM, Sophia-Antipolis, France\\
    		Email: \{flavio.maschietti, paul.dekerret, david.gesbert\}@eurecom.fr}
    }
\maketitle

\begin{abstract} 

	The synergetic gains of spectrum sharing and millimeter wave (mmWave) communication networks have recently attracted attention, owing to the 
	interference canceling benefits of highly-directional beamforming in such systems. In principle, fine-tuned coordinated scheduling and
	beamforming can drastically reduce cross-operator interference. However, this goes at the expense of the exchange of global channel state
	information, which is not realistic in particular when considering inter-operator coordination. Indeed, such an exchange of information is
	expensive in terms of backhaul infrastructure, and besides, it raises sensitive privacy issues between otherwise competing operators. 
	In this paper, we expose the existence of a trade-off between coordination and privacy. We propose an algorithm capable of balancing 
	spectrum sharing performance with privacy preservation based on the sharing of a low-rate beam-related information. Such information 
	is subject to a data obfuscation mechanism borrowed from the digital security literature so as to control the privacy, measured in terms
	of information-theoretical equivocation.
	
\end{abstract}

\section{Introduction} \label{sec:Intro}
	
	Millimeter wave communications ($30-300$ GHz) have given a renewed impetus to spectrum sharing, which allows multiple mobile operators 
	to pool their spectral resources. Compared to conventional (sub-$6$ GHz) mobile communications, less interference is in general produced in
	mmWave networks due to the inherent propagation characteristics and highly-directional beamforming~\cite{Heath2016, Samini2016}. 
	In particular, even without coordination, sharing spectrum and base stations (BSs) among operators shows great potential in mmWave scenarios 
	when massive antennas are used at both the BS and user (UE) sides~\cite{Gupta2016}. In addition to such technical gains, sharing resources 
	translates into substantial economic profit for the mobile operators. For example, dense infrastructure is an expected need for effective mmWave
	coverage in $5$G mobile networks and spectrum sharing among operators can help decrease equipment and operating costs~\cite{Rebato2016}. 
	In parallel, expenditure arising from spectrum licensing could be reduced as well.
	
	Although uncoordinated mmWave (shared) spectrum access is beneficial under certain circumstances, further gains can be achieved
	through inter-operator coordination. In particular, the gains are high when performance is understood not in a theoretically-relevant average
	throughput sense, but rather in a more practical reliable (or outage-constrained) throughput sense, i.e. looking at the near-worst case 
	scenarios. Indeed, catastrophic interference is experienced when e.g. non-massive antennas are used at the UE side, or also when the
	densities of either the UEs or the BSs increase, i.e. for reduced angular separation among the UEs~\cite{ShokriGhadikolaei2016} or 
	increased multi-cell interference.
	
	Nevertheless, the potential in coordinated spectrum sharing across operators implies several practical challenges. 
	For example, global CSI should be obtained for transmission optimization, leading to substantial signaling overhead. 
	Perhaps even more acute is the problem of \emph{data privacy preservation} between otherwise competing operators. 
	Since coordination entails some CSI flowing from one mobile operator to another, \emph{information privacy} issues emerge. 
	This problem is severe in mmWave networks where, owing to strong LOS propagation behavior~\cite{Samini2016}, CSI data
	bears correlation with UE location information, which for obvious reasons is undesirable for an operator to reveal~\cite{Liu2018}.
	
	In this work, we look at the \emph{trade-off} between coordination and privacy in mmWave spectrum sharing. 
	Up to our knowledge, this trade-off has not been investigated before. In line with other several mmWave studies~\cite{Maschietti2017, Ali2018}, 
	and in order to avoid severe overhead from CSI acquisition and exchange with massive antennas, we consider statistical side-information for
	transmission optimization. In particular, low-rate beam-related information is assumed to be exchanged between the operators.
	We propose then a low-overhead SLNR-based scheduling algorithm exploiting such information.
	To tackle the aforementioned privacy problem, we consider an information exchange scheme including a data obfuscation mechanism borrowed
	from the security literature~\cite{Duckham2005, Kido2005, Ardagna2007}. In mmWave spectrum sharing, this mechanism allows to 
	mitigate the one-to-one correspondence between beams and UEs' locations. The proposed algorithm manifests robustness towards the altered 
	side-information, and strikes a balance between average spectral efficiency (SE) and user \emph{confidentiality}.

\section{System Model and Problem Formulation}

	We consider a multi-cell multi-operator downlink mmWave scenario where $M$ mobile operators coexist and share the available mmWave spectrum.
	We consider $B$ BSs, all equipped with $N_{\text{BS}} \gg 1$ antennas, and $U$ associated UEs per BS, using single omnidirectional antennas. 
	To ease the exposition, we assume \emph{analog-only} beamforming with a single RF chain~\cite[Fig. $2$]{Heath2016}. Therefore, each BS uses a 
	single beam only per time-frequency resource slot. In particular, in a given slot, the $b$-th BS precodes the signal to the $u$-th UE using 
	the unit norm vector $\mathbf{w}_{b, u}$, extracted from a codebook with constant-magnitude elements, due to hardware constraints 
	(phase shifters)~\cite{Heath2016}.
	
	\begin{remark}	
		In general, mmWave communications exploit mixed analog-digital \emph{(hybrid)} precoding for reduced signal processing 
		and power consumption~\cite{Heath2016}. In this respect, $\mathbf{w}_{b, u} = \mathbf{w}_{b, u}^{\text{RF}} \mathbf{w}_{b, u}^{\text{D}}$,
		i.e. $\mathbf{w}_{b, u}$ results from the concatenation of a digital precoder with an analog (RF) one. 
		Here, $\mathbf{w}_{b, u}^{\text{D}} = 1$. \qed
	\end{remark}
	
	\subsection{Millimeter Wave Channel Model}
	
		Unlike the conventional sub-$6$ GHz band propagation environment, the mmWave one does not usually exhibit rich-scattering~\cite{Samini2016} 
		and can be modeled as a geometric channel with a limited number $L$ of dominant propagation paths. Therefore, the wideband mmWave channel 
		$\mathbf{h}_{b, u} \in \mathbb{C}^{N_{\text{BS}} \times 1}$ between the $b$-th BS and the $u$-th UE can be expressed as 
		follows~\cite{Ali2018}:
		\begin{equation} \label{H}
			\mathbf{h}_{b, u} \triangleq \sqrt{N_{\textrm{BS}}} \Big( \sum_{\ell=1}^L
				\alpha_{b, u, \ell} \mathbf{a}_{\textrm{BS}}(\theta_{b, u, \ell}, \phi_{b, u, \ell}) \Big)
		\end{equation}
		where $\alpha_{b, u, \ell} \sim \mathcal{CN}(0, \sigma_{\alpha_{b, u, \ell}}^2)$ denotes the complex gain of the $\ell$-th path -- 
		whose value includes the shaping filter (dependent on the path delay $\tau_{b, u, \ell}$) and the large-scale pathloss -- and where 
		$\mathbf{a}_{\textrm{BS}}(\theta_{b, u, \ell}, \phi_{b, u, \ell}) \in \mathbb{C}^{N_{\textrm{BS}} \times 1}$ denotes the antenna steering
		vector at the $b$-th BS with the corresponding \emph{angle-of-departure} (AoD) $(\theta_{b, u, \ell}, \phi_{b, u, \ell}) \in [0, 2\pi) 
		\times (0, \frac{\pi}{2}]$ in its azimuth and elevation components. In order to enable $3$D beamforming, we assume to use 
		\emph{uniform planar arrays} (UPA), so that~\cite{Heath2016}
		\begin{equation}
			\mathbf{a}_{\textrm{BS}}(\theta, \phi) \triangleq 
			\mathbf{a}_{\text{H}}(\theta, \phi) \otimes \mathbf{a}_{\text{E}}(\phi) 
		\end{equation}
		where $\otimes$ denotes the Kronecker product, and with
		\begin{equation}
			\mathbf{a}_{\text{H}}(\theta, \phi) \triangleq \sqrt{\frac{1}{N_{\text{BS}_{\textrm{H}}}}} \begin{bmatrix}
				1 & \dots & e^{-i \pi (N_{\text{BS}_{\textrm{H}}} - 1) \cos(\theta)\cos(\phi)}
				\end{bmatrix}^\mathrm{T},
		\end{equation}
		\begin{equation}
			\mathbf{a}_{\text{E}}(\phi) \triangleq \sqrt{\frac{1}{N_{\text{BS}_{\textrm{E}}}}} \begin{bmatrix}
				1 & \dots & e^{-i \pi (N_{\text{BS}_{\textrm{E}}} - 1) \sin(\phi)}
				\end{bmatrix}^\mathrm{T},
		\end{equation}
		where $N_{\text{BS}_{\textrm{H}}}$ (resp. $N_{\text{BS}_{\textrm{E}}}$) defines the number of the horizontal (resp. vertical) 
		UPA elements.
	
	\subsection{Analog Codebook}
	
		To design the beamforming vector $\mathbf{w}_{b, u}$, we assume -- as commmon in analog mmWave communications~\cite{Heath2016} -- 
		that each BS selects the beam configuration within a predefined beam codebook.
		To benefit from Full Dimensional (FD)-MIMO, a Discrete Fourier Transform (DFT)-based codebook has been proposed in~\cite{Xie2013}. 
		Such a codebook results from the Kronecker product of two oversampled DFT codebooks.
		In particular, we have
		\begin{equation} \label{DFT_Beams} 
			\mathbf{w}_{\eta(w, v)} \triangleq \mathbf{w}_{\text{H}, w} \otimes \mathbf{w}_{\text{E}, v}, 
			\quad w \in \llbracket 1, N_{\text{BS}_{\textrm{H}}} \rrbracket, 
			\quad v \in \llbracket 1, N_{\text{BS}_{\textrm{E}}} \rrbracket
		\end{equation}
		where $\mathbf{w}_{\text{H}, w}$ and $\mathbf{w}_{\text{E}, v}$ are as in~\cite[eq. (5)]{Xie2013},
		and $\eta(w, v) : \llbracket 1, N_{\text{BS}_{\textrm{H}}} 
		\rrbracket \times \llbracket 1, N_{\text{BS}_{\textrm{E}}} \rrbracket \rightarrow \llbracket 1, N_{\text{BS}} \rrbracket$ 
		is a bijection, e.g. $f(w, v) = N_{\text{BS}_{\textrm{E}}} (w-1) + v$, and $\llbracket 1, N_{\text{BS}} \rrbracket$ denotes the set 
		$\{ 1, \dots, N_{\text{BS}} \} \subset \mathbb{N}$.
		
	\subsection{Coordinated Time Division Scheduling Problem}
		
\begin{figure}[h]
			\centering
			\resizebox{15.7cm}{!}{
			\begin{tikzpicture}
				\draw (-6,0) rectangle (-3,1) node[pos=.5] {Time Slot $1$} node[pos=0.5, above of=0.5]{UEs $\{1, 8, 23\}$};
				\draw (-3,0) rectangle (0,1) node[pos=.5] {Time Slot $2$} node[pos=0.5, above of=0.5]{UEs $\{23, 11, 13\}$};
				\draw (0,0) rectangle (3,1) node[pos=.5] {$\dots$} node[pos=0.5, above of=0.5]{$\dots$};
				\draw (3,0) rectangle (6,1) node[pos=.5] {Time Slot $\smash{N_s - 1}$} node[pos=0.5, above of=0.5]{UEs $\{4, 7, 21\}$};
				\draw (6,0) rectangle (9,1) node[pos=.5] {Time Slot $\smash{N_s}$} node[pos=0.5, above of=0.5]{UEs $\{5, 17, 18\}$};
				\draw[|<->|] (-6,-0.78) -- (9, -0.78) node[pos=0.5,fill=white,inner sep=1pt]{Time Frame (Scheduling period)};
			\end{tikzpicture}
			}
			\caption{Time division scheduling with $B = 3$ and a sample assignment. In each time slot, each BS selects one UE to schedule.
			In this example, the BS $1$ chose the UEs $\{1, 23, \dots, 4, 5 \}$ overall.}
			\label{fig:time_division_sched}
		\end{figure}
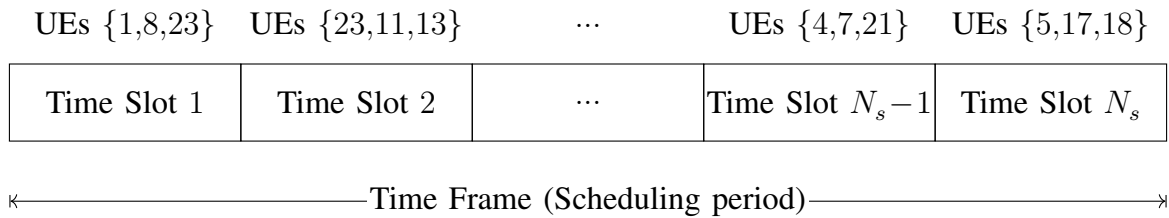

		We first present the centralized coordination problem towards spectrum sharing, based on scheduling and beamforming. 
		We assume a time division framework~\cite{Alizadeh2018} in which each scheduling period, i.e. a time frame with length $T$,
		is divided into $N_s$ slots with length $T_s = T / N_s$, as shown in Figure~\ref{fig:time_division_sched}. 
		The channel coherence time is assumed to be long enough so that all the UEs can be scheduled in one time frame. Based on their 
		available information, and aiming to improve performance, the BSs (belonging to different operators) assign one UE each per time slot.
		
		In the following, we assume that the association between BSs and UEs has been accomplished based on minimum UE-BS distance criterion. 
		The association between one BS and one UE in a mmWave network involves a beam choosing stage for which a transmit beam is selected to
		communicate~\cite{Alkhateeb2017}. We assume an SNR maximization scheme where the beam index 
		$\eta_u \in \llbracket 1, N_{\textnormal{BS}} \rrbracket$ chosen by the $b$-th BS to serve its $u$-th UE is as follows:
		\begin{equation}
			\eta_u = \argmax_{\eta \in \llbracket 1, N_{\textnormal{BS}} \rrbracket} ~|\mathbf{h}_{b, u} \mathbf{w}_{\eta}|^2.
		\end{equation}
		
		
		Let us denote with $\mathcal{S}(n)$ the set containing all the UEs scheduled in the time slot $n$. The instantaneous SINR for the 
		$u$-th UE, where $u \in \mathcal{S}(n)$, can be expressed as follows:
		\begin{equation} \label{SINR}
			\gamma_{u}(\mathcal{S}(n), \mathcal{P}) \triangleq \frac{\mathcal{P}_{u, u}}{
			\sum\limits_{q \in \mathcal{S}(n)} \mathcal{P}_{q, u} + \sigma^2_n}
		\end{equation}
		where we have defined the received power at the $u$-th UE being intended for the $q$-th one, as
		\begin{equation}
			\mathcal{P}_{q, u} \triangleq |\mathbf{h}_{j, u} \mathbf{w}_{\eta_q}|^2.
		\end{equation}
		\begin{remark} 
			We have made here the abuse of notation $\mathbf{h}_{j, u}$ to denote the channel between the $j$-th BS (associated with the $q$-th UE)
			and the $u$-th UE (associated with the $b$-th BS). The BS indexes $b$ and $j$ are thus implicit in $\mathcal{P}_{q, u}$ from now on. \qed
		\end{remark}
		
		The scheduling problem consists then in selecting the subset of UEs to schedule in each time slot so as to maximize the average network 
		sum-rate. Let $\mathcal{S} = \{\mathcal{S}(1), \dots, \mathcal{S}(N_s)\}$ denote the overall scheduling assignments, then the optimal
		scheduling decision $\mathcal{S}^*$ can be found as follows:
		\begin{equation} \label{Classic_Sched}
			\mathcal{S}^* = \argmax_{\mathcal{S}} \sum_{(u, n) \in \mathcal{S}(n) \times \llbracket 1, N_s \rrbracket}
			\log_2\big(1 + \gamma_{u}(\mathcal{S}(n), \mathcal{P})\big).
		\end{equation}
		
		The optimization problem in \eqref{Classic_Sched} is a challenging subset selection problem. In addition, to solve \eqref{Classic_Sched}, 
		the instantaneous CSI of all the UEs need to be shared across the BSs, or as an alternative be provided to a centralized coordinator.
		This requires unfeasible resource overhead. Moreover, in a spectrum sharing scenario, such information has to be shared and exchanged
		between different mobile operators. To ease overhead, we are interested instead in distributed approaches to solve the scheduling problem.
		In what follows, we first present a version of such algorithm \emph{without} privacy considerations, we then turn to the coordination-privacy
		trade-off in Section~\ref{sec:PP_Coo_Sched}.
		
\section{Greedy Successive Scheduling} \label{sec:Greedy_Sched}
	
	In the decentralized case, opposite to \eqref{Classic_Sched}, the operators need to enforce coordination while not being able to
	\emph{accurately} predict each other scheduling actions. Since each scheduling decision impacts on the overall network performance (and on
	the other scheduling decisions), the problem becomes even more challenging and requires some iterations with guessing. To go around this issue,
	we follow the well-known successive (or hierarchical) scheduling approach, such as presented in~\cite{Trivellato2007}.
	
	
	\subsection{SINR-Based Successive Coordinated Scheduling}
	
		In successive scheduling, a \emph{ranking} is first defined among the BSs and allows for consecutive scheduling decisions, 
		in a \emph{greedy sub-optimal} manner. In particular, at the $b$-th step of the successive scheduling algorithm, 
		the $b$-th BS knows the $b-1$ scheduling decisions made by the higher-ranked BSs $\{1, \dots, b-1\}$. In this work, we assume an
		\emph{arbitrary} ranking. Fixing some scheduling decisions allows to evaluate the so-called \emph{partial} SINR, in which the $b$-th BS
		solely considers the leakage coming from the UEs selected by the higher-ranked BSs in the considered time slot.
		Since the same operation is carried out for each time slot, we drop from now on the time slot index $n$ to lighten the notation. 
		Let us denote with $\mathcal{S}^b_{\textnormal{SINR}} = \{u^1_{\textnormal{SINR}}, \dots, u^b_{\textnormal{SINR}} \} = 
		\{\mathcal{S}^{b-1}_{\textnormal{SINR}}, u^b_{\textnormal{SINR}} \}$ the set consisting of all the scheduling decisions completed 
		at the $b$-th step of the successive scheduling. Then the \emph{partial} SINR $\hat{\gamma}_{u}$ for the $u$-th UE can be expressed 
		as follows:
		\begin{equation} \label{Partial_SINR}
			\hat{\gamma}_{u}(\mathcal{S}^{b-1}_{\text{SINR}},\mathcal{P}) \triangleq \frac{\mathcal{P}_{u, u}}
			{\sum\limits_{q \in \mathcal{S}^{b-1}_{\text{SINR}}}
			\mathcal{P}_{q, u} + \sigma^2_n}
		\end{equation}
		where the denominator includes the received power at the $u$-th UE being intended for the $q$-th one, where 
		$q \in \mathcal{S}^{b-1}_{\text{SINR}}$, i.e. the other UEs being scheduled in the considered time slot.
	
		Assuming that the scheduling information $\mathcal{S}^{b-1}_{\text{SINR}}$, from higher-ranked BSs $\{1, \dots, b-1\}$ have been 
		received\footnote{This information is assumed to be sent via dedicated channels and to be \emph{perfectly} decoded at the intended BS.} 
		at the $b$-th BS, the optimal successive scheduling decision $\mathcal{S}^b_{\text{SINR}}$ at the $b$-th BS can be expressed as follows:
		\begin{equation} \label{SINR_Succ_Sched}
			\mathcal{S}^b_{\text{SINR}} = \argmax_{u} ~\log_2\big(1 + \hat{\gamma}_{u}(\mathcal{S}^{b-1}_{\text{SINR}},\mathcal{P}) \big).
		\end{equation}
				
	\subsection{SLNR-Based Successive Coordinated Scheduling}
		
		Using the signal-to-leakage-and-noise ratio (SLNR) to optimize the scheduling decisions -- rather than the SINR as in
		\eqref{SINR_Succ_Sched} -- is advantageous as it does not require the knowledge of the channel between the considered $u$-th UE and 
		other BSs, which might belong to other operators. 
		Let us consider the $u$-th UE, then its \emph{partial} SLNR $\ubar{\gamma}_{u}$ can be expressed as follows:
		\begin{equation} \label{Partial_SLNR}
			\ubar{\gamma}_{u}(\mathcal{S}^{b-1}_{\text{SLNR}},\mathcal{P}) \triangleq \frac{\mathcal{P}_{u, u}} 
			{\sum\limits_{q \in \mathcal{S}^{b-1}_{\text{SLNR}}}
			\mathcal{P}_{u, q} + \sigma^2_n}
		\end{equation} 
		where, as opposite to \eqref{Partial_SINR}, the denominator includes the leakage $\mathcal{P}_{u, q}$ produced by the $u$-th UE on the 
		other UEs being scheduled in the considered time slot, denoted with $\mathcal{S}^{b-1}_{\text{SLNR}}$.
		
		Assuming that the scheduling information $\mathcal{S}^{b-1}_{\text{SLNR}}$ from higher-ranked BSs $\{1, \dots, b-1\}$ have been received, 
		the optimal SLNR-based successive scheduling decision $\mathcal{S}^b_{\text{SLNR}}$ at the $b$-th BS is obtained through solving the
		following optimization problem:
		\begin{equation} \label{SLNR_Succ_Sched}
			\mathcal{S}^b_{\text{SLNR}} = \argmax_{u} ~\ubar{\gamma}_{u}(\mathcal{S}^{b-1}_{\text{SLNR}},\mathcal{P}).
		\end{equation}
		
		Note that the above requires instantaneous CSI in principle. However, the method can be modified to leverage statistical CSI 
		instead as is shown below.
		
	\subsection{Average Leakage Power Through Beam Footprints} \label{sec:Avg_Leakage}
		
		To reduce the severe overhead arising from global CSI exchange with massive antennas, we seek a coordination protocol which instead 
		allows exchanging low-rate\footnote{The so-called \emph{beam coherence time} has been reported to be in general much longer than the channel
		coherence time~\cite{Va2017}.} \emph{beam index} information between the operators. In the following, we show that such 
		information allows the BSs to estimate the potential (average) SLNR, without resorting to instantaneous CSI.
		In order to achieve this, we assume that when the $b$-th BS receives the scheduling information $\mathcal{S}^{b-1}$, a beam-related
		information $\eta_q, q \in \mathcal{S}^{b-1}$ is appended as well by the higher ranked BSs.	
		
		Let us consider the leakage $\mathcal{P}_{u, q}$ for a full-LOS case, i.e. $\alpha_{b, u, \ell}^2 = 0~\forall \ell$ relative to NLOS paths.
		We are interested in its expected value (over small-scale fading), which is
		\begin{align} \label{Perf_Avg_Rec_Power}
			\mathbb{E}\big[\mathcal{P}_{u, q}\big] &= \mathbb{E}_{\alpha_{b, q}}\big[
			|\sqrt{N_{\textrm{BS}}} \alpha_{b, q} \mathbf{a}_{\textrm{BS}}(\theta_{b, q}, \phi_{b, q}) \mathbf{w}_{\eta_u}
			|^2\big] \nonumber \\
			&= \mathbb{E}_{\alpha_{b, q}}\big[ |\mathcal{G}_{\eta_u}(\theta_{b, q}, \phi_{b, q}) \alpha_{b, q}|^2\big] \nonumber \\
			&= \mathcal{G}_{\eta_u}(\theta_{b, q}, \phi_{b, q}) \sigma^2_{\alpha_{b, q}}
		\end{align}
		where $\mathcal{G}_{\eta_u}(\theta_{b, q}, \phi_{b, q})$ denotes the beamforming gain received at the $q$-th UE with the beam intended 
		for the $u$-th one. 
		
\begin{figure}[h]
			\centering
			\includegraphics[trim=4cm 3.71in 4cm 10.21cm, scale=0.57]{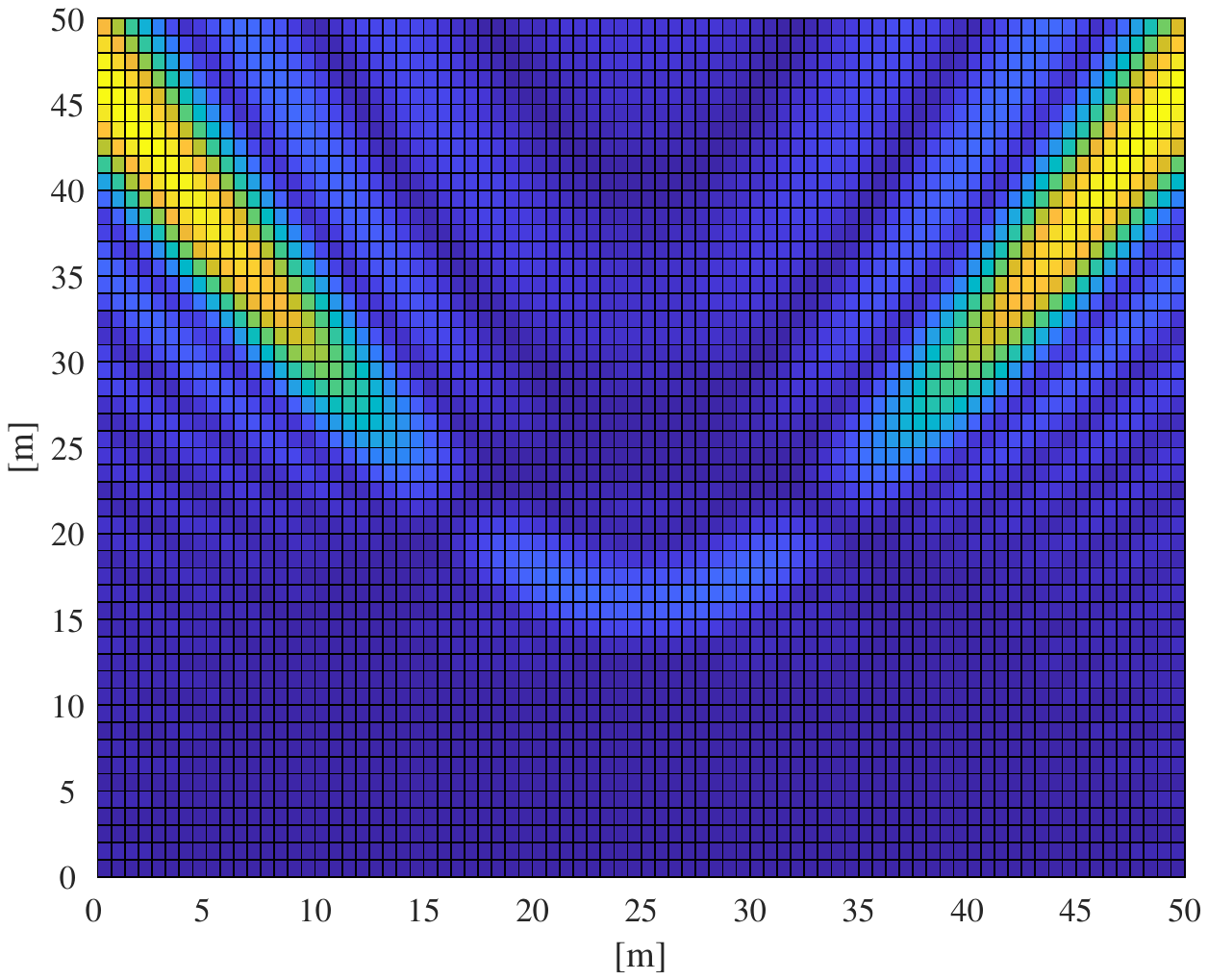}
			\quad
			\includegraphics[trim=4cm 3.71in 4cm 10.21cm, scale=0.57]{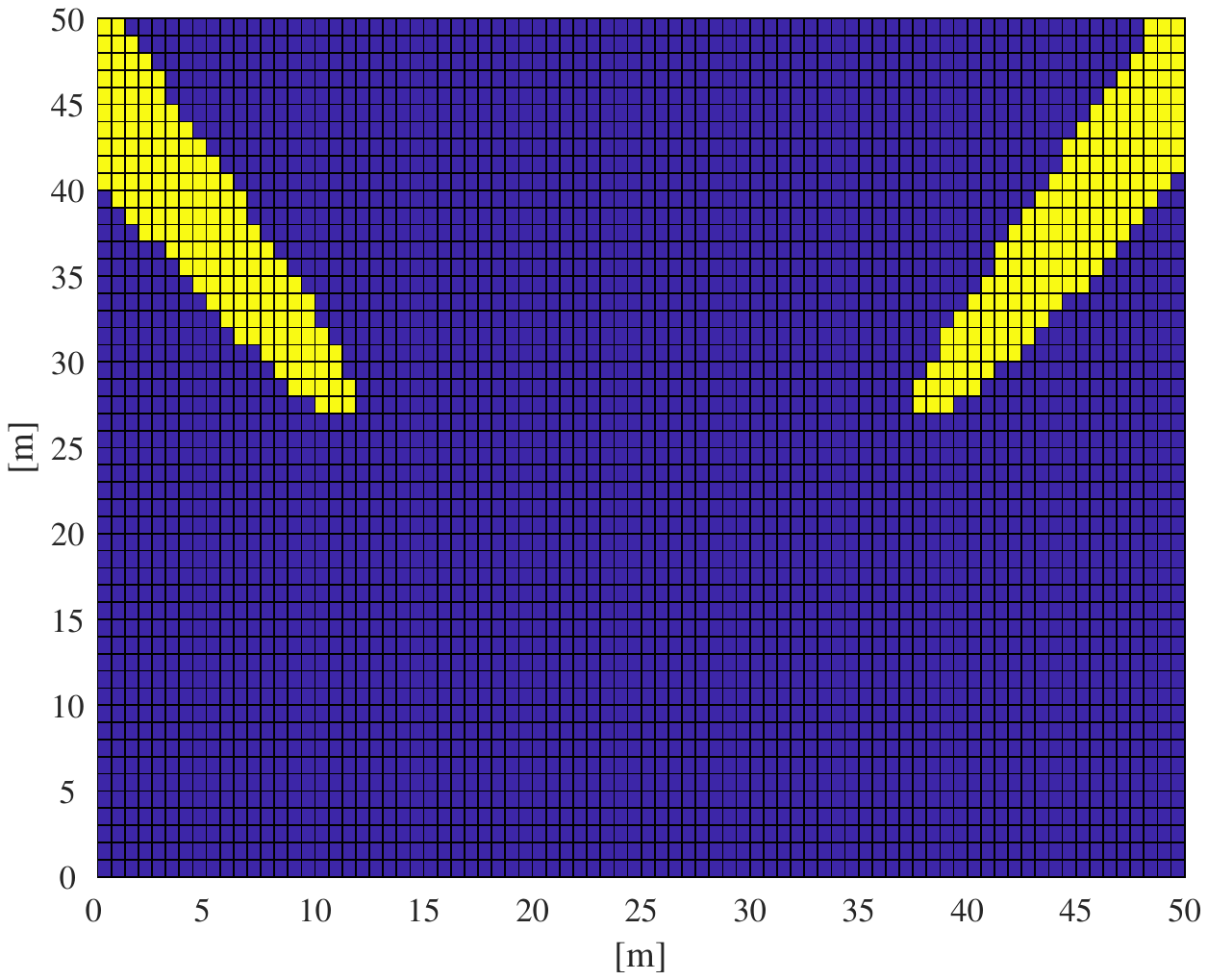}
			\caption{Beamforming gain per location obtained with two beams in \eqref{DFT_Beams} and their associated footprints, 
			considered as the spatial region where the normalized gain is higher than $1/2$.}
			\label{fig:beam_footprints}
		\end{figure}

		To evaluate \eqref{Perf_Avg_Rec_Power}, the $b$-th BS needs to know the AoD $(\theta_{b, q}, \phi_{b, q})$ and the 
		average path gain $\sigma^2_{\alpha_{b, q}}$. Note that, although the latter is a long-term \emph{locally-available} statistical information
		(it is the average gain observed on a particular local direction), the former is hard to obtain in a scenario with multiple operators. 
		Still, beam-related information exchanged with the $j$-th BS can assist in evaluating $\mathbb{E}\big[\mathcal{P}_{u, q}\big]$. 
		In particular, the beams in \eqref{DFT_Beams} concentrate on different spatial regions~\cite{Xie2013}. In practice, their main lobes
		illuminate non-overlapping regions, also known as \emph{beam footprints} (refer to Fig.~\ref{fig:beam_footprints}).
		
		As a consequence, beam-related information might \emph{implicitly} circumscribe the UEs' locations within the \emph{beam footprints} -- 
		in particular in LOS-dominated environments as the mmWave one~\cite{Samini2016}. Let us assume that the $q$-th UE is served through a LOS
		path, then we can bound its actual location $\boldsymbol{\ell}_q \in \mathbb{R}^2$ within the footprint of its serving beam $\eta_q$.
		It is possible then to compute the average leakage $\mathbb{E}\big[\mathcal{P}_{u, q}\big]$ with respect to all the plausible positions 
		of the $q$-th UE within the footprint of $\eta_q$. In particular, we can evaluate $\mathbb{E}\big[\mathcal{P}_{u, q}\big]$ as follows:
		\begin{align} \label{Avg_Leakage}
			\mathbb{E}\big[\mathcal{P}_{u, q}\big] &= 
			\mathbb{E}_{(\theta_{b, q}, \phi_{b, q})|\eta_u} 
			\big[\mathcal{G}_{\eta_u}(\theta_{b, q}, \phi_{b, q}) \sigma^2_{\alpha_{b, q}} \big] \nonumber \\
			&= \int_{(\theta_{b, q}, \phi_{b, q}) \in \mathcal{Q}_{\eta_q}} 
			\mathcal{G}_{\eta_u}(\theta_{b, q}, \phi_{b, q}) 
			\sigma^2_{\alpha_{b, q}} d(\theta_{b, q}, \phi_{b, q}) \nonumber \\
			&\stackrel{(a)}{=} \int_{(\theta_{b, q}, \phi_{b, q}) \in \mathcal{Q}_{\eta_q} \cap \mathcal{Q}_{\eta_u}} G
			\sigma^2_{\alpha_{b, q}} d(\theta_{b, q}, \phi_{b, q}) 
			+ \int_{(\theta_{b, q}, \phi_{b, q}) \notin \mathcal{Q}_{\eta_q} \cap \mathcal{Q}_{\eta_u}} g
			\sigma^2_{\alpha_{b, q}} d(\theta_{b, q}, \phi_{b, q})
		\end{align}
		where $\mathcal{Q}_\eta$ contains the AoDs related to the footprint of the generic beam $\eta \in \llbracket 1, N_{\text{BS}} \rrbracket$,
		and where $(a)$ follows the well-known sectored antenna model~\cite{Andrews2017}, i.e.
		\begin{equation}
			\mathcal{G}_{\eta}(\theta, \phi) \triangleq \begin{cases}
				G, & ~(\theta, \phi) \in \mathcal{Q}_{\eta} \\
				g, & ~\text{otherwise}
			\end{cases}
		\end{equation}
		which results in considering $\mathcal{G}_{\eta_q}(\theta_{j, u}, \phi_{j, u}) = G$ in the overlapping sector of the footprints 
		relative to the $u$-th and the $q$-th UEs, and $\mathcal{G}_{\eta_q}(\theta_{j, u}, \phi_{j, u}) = g$ in the non-overlapping one.
	
	\subsection{Low-Overhead SLNR-Based Coordinated Scheduling}
		
		In this section, we introduce the proposed low-overhead SLNR-based scheduling algorithm exploiting the beam-related information 
		(as described in Section \ref{sec:Avg_Leakage}) available at each operator. The intuition behind such an approach is that the UEs served
		with beams whose footprints are non-overlapping (or partially-overlapping) can be scheduled \emph{simultaneously}, aiming to reduce the
		overall interference and maximize the network SE.
		
		Let us denote with $\mathcal{S}^{b-1}_{\text{LOW}}$ the scheduling information -- here including both scheduling ordering and 
		appended beam-related information -- received from higher-ranked BSs $\{1, \dots, b-1\}$. Then, the scheduling decision 
		$\mathcal{S}^b_{\text{LOW}}$ at the $b$-th BS can be obtained as follows:
		\begin{equation} \label{SLNR_Low_Sched_Simple}
			\mathcal{S}^b_{\text{LOW}} = \argmax_{u} ~\bar{\gamma}_{u}(\mathcal{S}^{b-1}_{\text{LOW}},\hat{\mathcal{P}}^b_\text{LOW}) 
		\end{equation}
		where $\bar{\gamma}_{u}$ is the approximated average \emph{partial} SLNR defined as 
		\begin{equation} \label{Avg_Partial_SLNR}
			\bar{\gamma}_{u}(\mathcal{S}^b_{\text{LOW}},\mathcal{P}) \triangleq \frac{\mathbb{E}\big[\mathcal{P}_{u, u}\big]}
			{\sum\limits_{q \in \mathcal{S}^b_{\text{LOW}}} \mathbb{E}\big[\mathcal{P}_{u, q}\big] + \sigma^2_n}
		\end{equation}
		and where $\hat{\mathcal{P}}^b_\text{LOW}$ collects all the required $\mathbb{E}\big[\mathcal{P}_{u, q}\big] ~\forall q
		\in \mathcal{S}^{b-1}_{\text{LOW}}$ at the $b$-th BS, estimated through \eqref{Avg_Leakage}.
		
		\begin{remark}
			The computation of the required $\hat{\mathcal{P}}^b_\text{LOW}$ can be done once for a given scenario as it depends solely on the 
			beam footprints, which are static for some fixed cooperating BSs. \qed
		\end{remark}
		
		We summarize the proposed low-overhead SLNR-based coordinated scheduling algorithm in Algorithm~\ref{PseudoCode}.
		The average leakage in \eqref{Avg_Leakage} is evaluated through numerical integration.
		
		\makeatletter \xpatchcmd{\algorithmic}{\itemsep\z@}{\itemsep=0.71ex}{}{}\makeatother
		\algrenewcommand\algorithmicindent{0.71em}
		\begin{algorithm} 
			\caption{Low-Overhead SLNR-Based Coordinated Successive Scheduling at the $b$-th BS for a given time slot} 
			\label{PseudoCode}
			\begin{algorithmic}[1]
				\small
				\Statex INPUT: $\mathcal{S}^{b-1}_{\text{LOW}}$, $\eta_u ~\forall u \in \llbracket 1, U \rrbracket$, 
				$\hat{\mathcal{P}}^b_{\text{LOW}}$
					\If{$b = 1$} \Comment{The $b$-th BS is the first to decide}
						\State $\mathcal{S}^b_{\text{LOW}} \leftarrow \argmax_{u} |\mathbf{h}_{b, u} \mathbf{w}_{\eta_u}|^2$ 
						\Comment {SNR-based scheduling}
					\Else \Comment{The $b$-th BS is not the first to decide}
						\State Retrieve $\mathbb{E}\big[\mathcal{P}_{u, q}\big] 
						~\forall q \in \mathcal{S}^{b-1}_{\text{LOW}}$ from $\hat{\mathcal{P}}^b_\text{LOW}$
						\State $\mathcal{S}^b_{\text{LOW}} \leftarrow$ Solve \eqref{SLNR_Low_Sched_Simple} using the retrieved information
					\EndIf
					\State \textbf{return} $\mathcal{S}^b_{\text{LOW}}$
			\end{algorithmic}
		\end{algorithm} \vspace{-0.11cm}
		
\section{Privacy-Preserving Coordinated Scheduling} \label{sec:PP_Coo_Sched}
		
		In the previous section, we have introduced a low-overhead scheduling algorithm exploiting beam-related information. In particular, 
		such approach relies on estimating the leakage through beam footprints. In this section, aware of the \emph{information privacy} issues
		outlined in Section \ref{sec:Intro}, we propose a \emph{privacy-preserving} exchange mechanism allowing coordination between the
		operators. Then, we introduce a robust scheduling algorithm exploiting the altered beam-related information.	
	
	\subsection{Trade-Off Between Coordination and Privacy} \label{sec:Beam_Info}
		
		As described in Section~\ref{sec:Avg_Leakage}, beam-related information might \emph{implicitly} offer an insight into the UEs' locations.
		If the $u$-th UE is served through a LOS path, then we can bound its actual location $\boldsymbol{\ell}_q \in \mathbb{R}^2$
		within the footprint of its serving beam $\eta_q$. In particular, assuming \emph{uniformly}-distributed UEs in the network area 
		$\mathcal{A}$, we can write the PDF $f(\boldsymbol{\ell}_q | \eta_q)$ as follows:
		\begin{equation} \label{Pos_PDF}
			f(\boldsymbol{\ell}_q | \eta_q) \triangleq \begin{cases}
				0, & \boldsymbol{\ell}_q \notin \mathcal{A}_{\eta_q} \subset \mathcal{A} \\
				|\mathcal{A}_{\eta_q}|^{-1}, & \boldsymbol{\ell}_q \in \mathcal{A}_{\eta_q} \subset \mathcal{A}
			\end{cases}
		\end{equation}
		where $\mathcal{A}_{\eta_q}$ is the footprint relative to $\eta_q$, and $|\mathcal{A}_{\eta_q}|$ is its area.
		
		We are interested in evaluating how uncertain is the generic BS about $\boldsymbol{\ell}_q$ given $\eta_q$. 
		This can be measured through the information-theoretical equivocation, which also indicates the \emph{confidentiality}
		attributed to $\boldsymbol{\ell}_q$~\cite{Wagner2018}. The equivocation is defined conventionally as follows:
		\begin{align} \label{Equivocation_SameOp}
			H(\boldsymbol{\ell}_q | \eta_q)
			&\triangleq - \int_{\boldsymbol{\ell}_q \in \mathcal{A}_{\eta_q}} 
			f(\boldsymbol{\ell}_q | \eta_q) \log_2(f(\boldsymbol{\ell}_q | \eta_q)) d\boldsymbol{\ell}_q \nonumber \\
			&= \log_2(|\mathcal{A}_{\eta_q}|).
		\end{align}
		
		Sending \emph{obfuscated} beam-related information to other operators involves injecting on purpose some additional \emph{uncertainty} 
		about the actual location $\boldsymbol{\ell}_q \in \mathbb{R}^2$ of the $q$-th UE. 
		In this respect, an operator can provide increased \emph{privacy} to its customers.
		Spatial information is in general obfuscated through enhancing its \emph{inaccuracy}, i.e. the incorrespondence between information 
		and actual location, and \emph{imprecision}, i.e. the inherent vagueness in location information~\cite{Duckham2005, Kido2005, Ardagna2007}. 
		For example, in~\cite{Kido2005}, several false locations (dummies) are associated to each protected and real UE, thus making its location
		information harder to infer. We consider an equivalent obfuscation mechanism for which multiple possible beams (thus locations) are
		associated to the $q$-th UE. Let $\boldsymbol{\eta}_q^b$ denote the information about $\eta_q$ available at the $b$-th BS. 
		Considering for the sake of exposition that each BS belongs to a different operator, we have
		\begin{equation} \label{Ob_Beam}
			\boldsymbol{\eta}_q^b = \{ \eta_{\omega_q(1)}, \dots, \eta_{\omega_q(K)}, \eta_q \}
		\end{equation}
		where $\omega_q : \llbracket 1, K \rrbracket \rightarrow \llbracket 1, N_{\text{BS}} \rrbracket$ is the deterministic obfuscating
		function relative to the $q$-th UE, with $K$ being the number of obfuscating beams (or dummy beams). 
		
		\begin{lemma}
			Following the obfuscation mechanism, the equivocation on $\boldsymbol{\ell}_q$ can be expressed as follows:
			\begin{align}
				H(\boldsymbol{\ell}_q | \boldsymbol{\eta}_q^b)
				& = - \sum_{\eta \in \boldsymbol{\eta}_q^b} \int_{\boldsymbol{\ell}_q \in \mathcal{A}}
				f(\boldsymbol{\ell}_q, \eta) \log_2(f(\boldsymbol{\ell}_q | \eta)) d\boldsymbol{\ell}_q \nonumber \\
				&=- \sum_{\eta \in \boldsymbol{\eta}_q^b}
				\frac{1}{(K+1)} \int_{\boldsymbol{\ell}_q \in \mathcal{A}_{\eta}} 
				f(\boldsymbol{\ell} | \eta) \log_2(f(\boldsymbol{\ell}_q | \eta)) d\boldsymbol{\ell}_q \nonumber \\
				&= - \sum_{\eta \in \boldsymbol{\eta}_q^b} \frac{1}{(K+1)} 
				\int_{\boldsymbol{\ell}_q \in \mathcal{A}_{\eta}} 
				\frac{-\log_2 \big((K+1)|\mathcal{A}|\big)}{(K+1)|\mathcal{A}|}
				 d\boldsymbol{\ell}_q \nonumber \\
				&=- \sum_{\eta \in \boldsymbol{\eta}_q^b}
				\frac{-\log_2\big((K+1)|\mathcal{A}|\big)}{K + 1} \nonumber \\
				&= \log_2\big((K+1)|\mathcal{A}_{\eta_q}|\big)
			\end{align}
			where we have assumed that the area illuminated with the beams in $\boldsymbol{\eta}_q^b$ is the same\footnote{Although the beams in
			\eqref{DFT_Beams} illuminate bigger regions as the elevation angle increases, the UEs are expected to reside on average within regions
			($30^\circ-60^\circ$ in elevation) where the beam footprints can be assumed to be almost identical.} as $|\mathcal{A}_{\eta_q}|$.
		\end{lemma}
		
		The obfuscation mechanism results in a $\log_2(K+1)$ factor added to the equivocation in
		\eqref{Equivocation_SameOp} obtained with non-obfuscated information $\eta_q$, i.e. exchanges between the same operator.
	
	\subsection{Privacy-Preserving SLNR-Based Coordinated Scheduling}
		
		In a robust scheduling decision, each operator should account for the alterations in the exchanged beam-related information. 
		In practice, the expectation in \eqref{Avg_Leakage} needs to be further averaged over all the possible footprints to which the $q$-th UE
		might belong to. In order to avoid dealing with the expectation -- which could be approximated (with a discrete summation) through 
		Monte-Carlo iterations -- we consider the following conservative approach leading to a much less complex algorithm.
		
		Let us consider the obfuscated and received beam-related information $\boldsymbol{\eta}_q^b$. 
		Given such information, the $b$-th BS knows the set of the plausible beams used to serve the $q$-th UE. 
		In order to derive a simple scheduling decision, the $b$-th BS can assume that all those beams are \emph{actually} being used to 
		serve some \emph{phantom} UEs, and evaluate their average leakage through \eqref{Avg_Leakage}.
		
		Let us denote with $\mathcal{S}^{b-1}_{\text{ROB}}$ the scheduling information -- here enlarged with spurious obfuscating information --
		received from higher-ranked BSs $\{1, \dots, b-1\}$. Then, the robust \emph{privacy-preserving} scheduling decision 
		$\mathcal{S}^b_{\text{ROB}}$ at the $b$-th BS is obtained as follows:
		\begin{equation} \label{SLNR_Rob_Sched_Simple}
			\mathcal{S}^b_{\text{ROB}} = \argmax_{u} ~\bar{\gamma}_{u}(\mathcal{S}^{b-1}_{\text{ROB}},\hat{\mathcal{P}}^b_\text{ROB}) 
		\end{equation}
		where $\bar{\gamma}_{u}$ is the approximated \emph{partial} SLNR defined in \eqref{Avg_Partial_SLNR}.
		
		The robust scheduling algorithm can be solved via the proposed low-overhead Algorithm~\ref{PseudoCode}, 
		substituting $\mathcal{S}^{b-1}_{\text{LOW}}$ and $\hat{\mathcal{P}}^b_\text{LOW}$ with the enlarged 
		$\mathcal{S}^{b-1}_{\text{ROB}}$ and $\hat{\mathcal{P}}^b_\text{ROB}$, respectively.
		
		\begin{remark}
			Solving the optimization in \eqref{SLNR_Rob_Sched_Simple} means considering the alterations in the exchanged information, 
			but not the fact that the UEs in $\mathcal{S}^{b-1}_{\text{ROB}}$ might not be in LOS with their associated BSs.
			In mmWave networks, the percentage of NLOS links is small~\cite{Samini2016}. Still, a performance loss due to mismatches 
			is expected, and will be quantified in the following Section~\ref{sec:Sims}. \qed
		\end{remark}

\section{Simulation Results} \label{sec:Sims}

	We evaluate here the performance of the proposed scheduling algorithms. We assume that the BSs are non-colocated (no infrastructure sharing
	between the operators) and equipped with $N_{\text{BS}} = 128$ antennas ($16 \times 8$ UPA). We start with a simple non-dense scenario with
	$M = 2$ mobile operators and $B = 2$ BSs, one each operator. We assume a squared network area with side equal to $50$ m. We further assume
	$U = 10$ UEs per BS/operator and $N_s = 10$ scheduling time slots in which the channel is assumed to be coherent. All the plotted data rates are
	the averaged -- over $10^5$ Monte-Carlo runs -- instantaneous rates.
	
	\subsection{Results and Discussion}
		
		We consider stronger (on average) LOS paths with respect to the NLOS ones~\cite{Samini2016}. 
		In particular, we adopt the following large-scale pathloss model:
		\begin{equation}
			\text{PL}(\delta) = \alpha + \beta \log_{10}(\delta) + \xi \qquad [\text{dB}]
		\end{equation}
		where $\delta$ is the path length and the parameters $\alpha$, $\beta$, $\xi$ are taken from Tables III and IV 
		in~\cite{Samini2016} for both LOS and NLOS paths. 
		
		We introduce now the average UE \emph{detection probability} (DP) so as to relate the information-theoretical equivocation to a physical
		privacy metric. Intuitively, the DP measures the likelihood to correctly infer the location of the UEs -- up to a given area $X$ -- from 
		the exchanged information. It is defined as
		\begin{equation}
			\text{DP} = \mathbb{E}_q \left[ \frac{X}{(K+1)|\mathcal{A}_{\eta_q}|} \right].
		\end{equation}
		
\begin{figure}[h]
			\centering
			\includegraphics[trim=4cm 3.71in 4cm 10cm, scale=1.07]{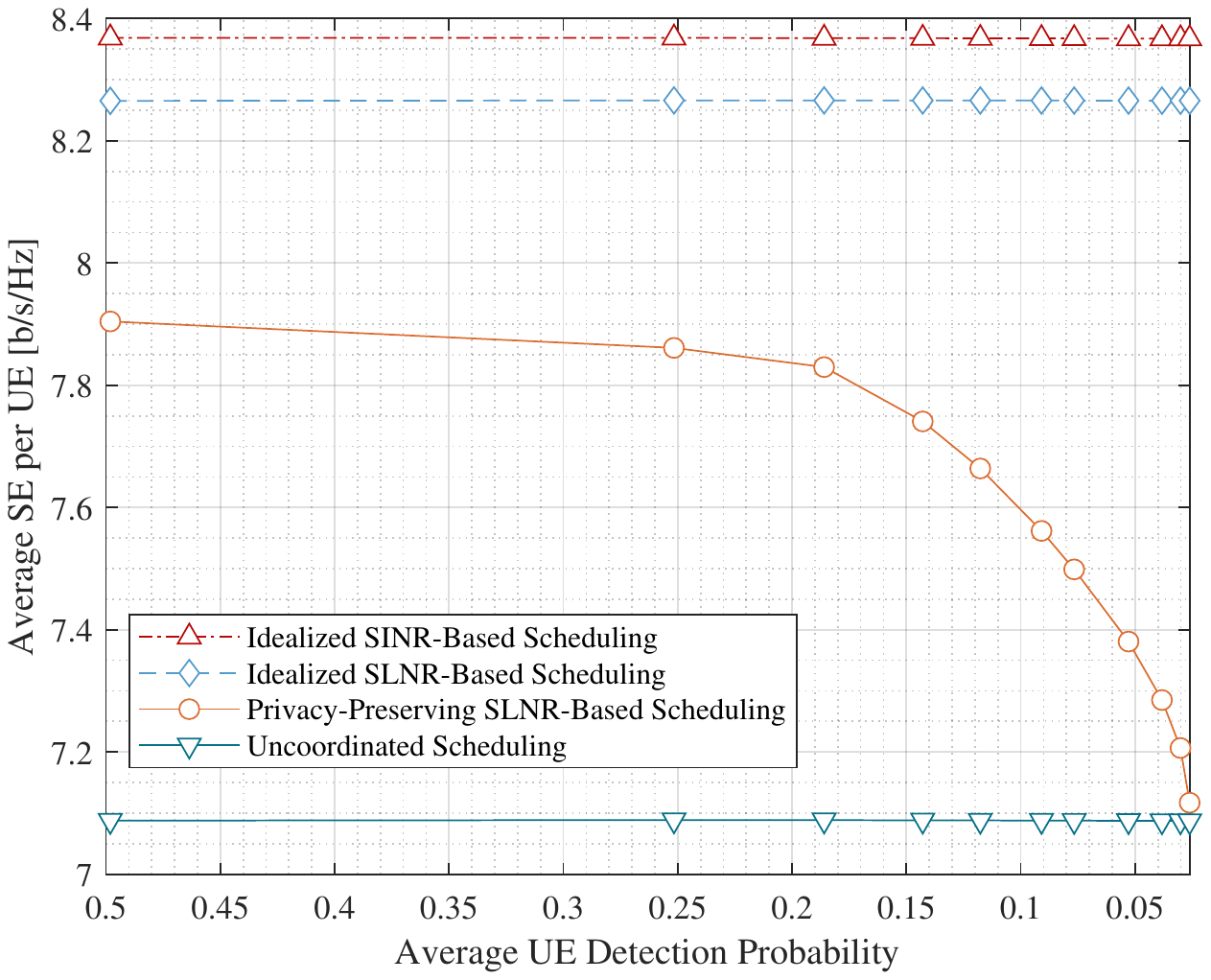}
			\caption{Average SE per UE vs Average DP in a full-LOS scenario.
			The proposed \emph{privacy-preserving} algorithm succeeds in striking a balance between privacy and average SE performance.}
			\label{fig:avg_SE_vs_Eqvc}
		\end{figure}

		In Fig.~\ref{fig:avg_SE_vs_Eqvc}, we show the performance of the proposed algorithm as a function of the UE \emph{detection probability}, 
		in a full-LOS scenario, i.e. $\alpha_{b, u, \ell}^2 = 0~\forall \ell$ relative to NLOS paths. The UE DP is controlled through the number of dummy
		beams $K$ in the exchanged information. Note that the parameter $K$ impacts our proposed \emph{privacy-preserving} algorithm only.
		The idealized scheduling algorithms and the uncoordinated one have a fixed DP level, which is $\mathbb{E}\big[ X / |\mathcal{A}_{\eta_q}|\big]$.
		
		In~\cite{Kido2005}, two algorithms have been proposed so as to generate \emph{realistic} false locations, which should exhibit some correlation
		with the actual location data. We generate instead the dummy beams according to a discrete uniform distribution over 
		$\llbracket 1, N_{\text{BS}} \rrbracket$ for simplicity, and consider their obfuscating properties as in a one-shot exchange mechanism.
		
		Note that even with $K = 0$ (no dummy beams), there is still a remaining \emph{uncertainty} with respect to the UEs' location, as the UEs
	 	can reside anywhere within their beam footprints, in this case larger than $X = 10$ m$^2$. The gap for $K = 0$ between the proposed
		coordinated algorithm and the idealized ones -- obtained with perfect knowledge of the matrix $\mathcal{P}$ -- is due to both average 
		SLNR and sectored antennas approximations. Our \emph{privacy-preserving} scheduling algorithm converges to 
		the uncoordinated solution (based on SNR, i.e. neglecting interference) as the average DP decreases, i.e. \emph{higher privacy}. 

\begin{figure}[h]
			\centering
			\includegraphics[trim=4cm 3.71in 4cm 10cm, scale = 1.07]{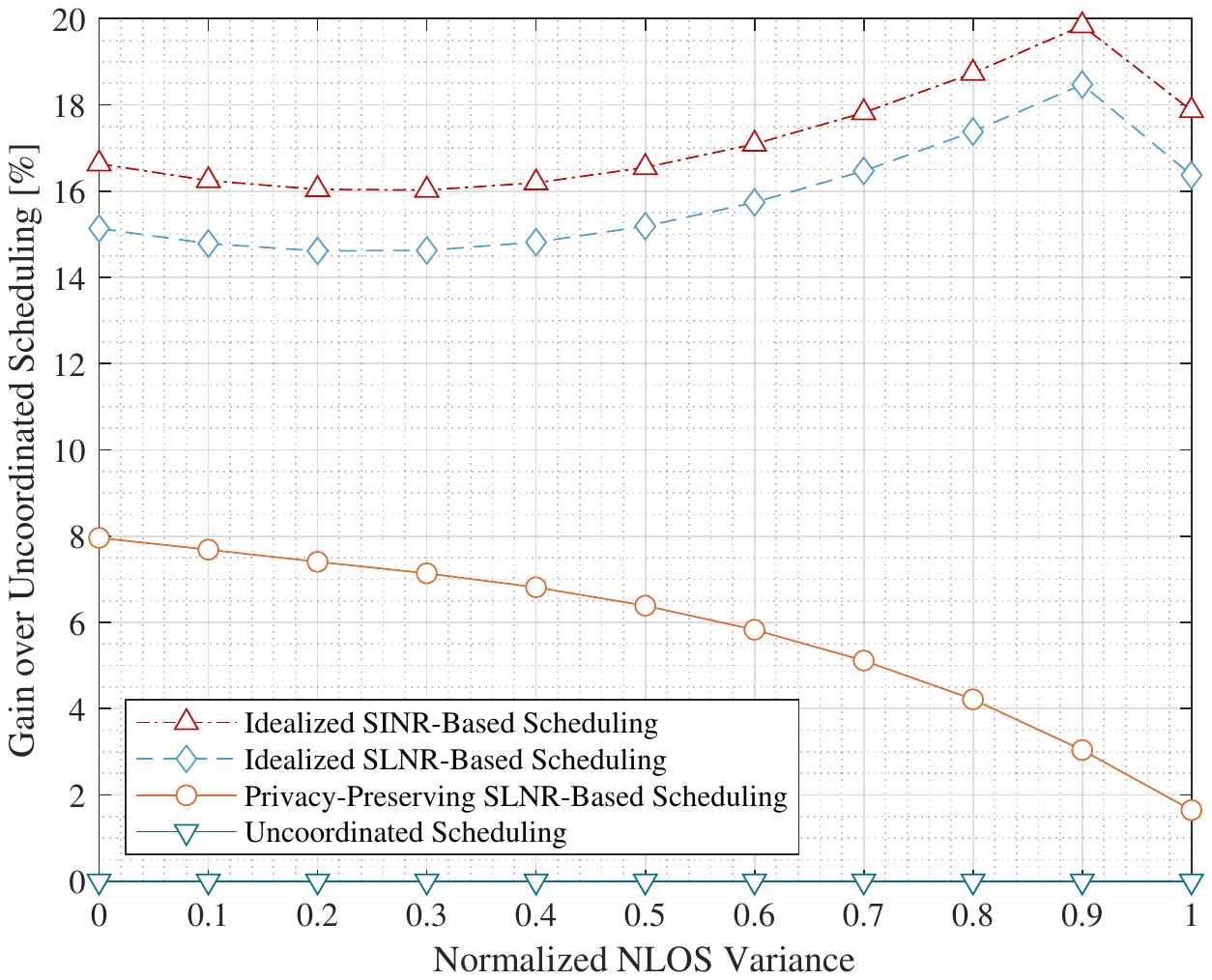}
			\caption{Gain over uncoordinated scheduling vs Normalized NLOS variance. Here, the UE $\text{DP} \simeq 0.1$. The performance 
			of the proposed \emph{privacy-preserving} low-overhead scheduling algorithm decreases as more NLOS links are used to communicate.}
			\label{fig:avg_SE_vs_ProbNLOS}
		\end{figure}

		In Fig.~\ref{fig:avg_SE_vs_ProbNLOS}, we measure the performance loss due to the NLOS/LOS mismatch, for a given DP, with $L = 5$ paths. 
		In this plot, we assume $\sum_{\ell} \hat{\sigma}^2_{b, u, \ell} = 1 ~\forall b, u$, where $\hat{\sigma}^2_{b, u, \ell}$ is the normalized
		variance of the $\ell$-path of $\mathbf{h}_{b, u}$. As expected, the proposed low-overhead coordinated algorithm loses up to a $7 \%$ over
		the uncoordinated solution as the variance of the NLOS links increases, which means that more NLOS paths are chosen as best 
		path for communicating. There still exists a gap between the proposed algorithm and the uncoordinated one for a full-NLOS scenario. 
		Indeed, the knowledge of the pathloss is exploited in the proposed algorithm, for which UEs which are quite far from each other
		are preferred for simultaneous scheduling.
		
\section{Conclusion}

	Dealing with inter-operator interference in mmWave spectrum sharing is essential for improving performance. Since multiple mobile operators
	are involved in the operation, \emph{privacy-preserving} mechanisms and distributed approaches to performance maximization are suitable. 
	In this work, we have proposed a low-overhead distributed SLNR-based scheduling algorithm exploiting obfuscated beam-related side-information.
	Numerical results indicate that a substantial gain is achieved through inter-operator cooperation even in non-dense scenarios with few
	operators/BSs. Further performance gain is expected in richer scenarios.
	
\section{Acknowledgments}

The authors are supported by the ERC under the European Union's Horizon 2020 research and innovation program (Agreement no. 670896 PERFUME).

\bibliography{Bibl}
\bibliographystyle{IEEEtran}
				
\end{document}